\documentclass[aoas,MSNbibl,nameyear,dvips]{arximspdf}
\usepackage{dcolumn}
\usepackage{graphicx}

\doi{10.1214/13-AOAS713} 
\volume{8}
\issue{1}
\pubyear{2014}
\firstpage{204}
\lastpage{231}

\makeatletter
\newcolumntype{d}[1]{D{.}{.}{#1}}
\newcommand{\rrvert}{\vert}
\newcommand{\llvert}{\vert}

\makeatother

\begin{document}
\begin{frontmatter}

\title{Matching for balance, pairing for heterogeneity in an
observational study of
the effectiveness of for-profit and not-for-profit high schools in Chile}
\runtitle{Matching for balance, pairing for heterogeneity}

\pdftitle{Matching for balance, pairing for heterogeneity in an
observational study of
the effectiveness of for-profit and not-for-profit high schools in Chile}

\begin{aug}
\author[A]{\fnms{Jos\'{e} R.} \snm{Zubizarreta}\corref{}\ead[label=e1]{zubizarreta@columbia.edu}},
\author[B]{\fnms{Ricardo D.} \snm{Paredes}\ead[label=e2]{rparedes@ing.puc.cl}}\\
\and
\author[C]{\fnms{Paul R.} \snm{Rosenbaum}\ead[label=e3]{rosenbaum@wharton.upenn.edu}\thanksref{t1}}
\runauthor{J. R. Zubizarreta, R. D. Paredes and P. R. Rosenbaum}
\affiliation{Columbia University, Pontificia Universidad Cat\'{o}lica de Chile\\ and University of Pennsylvania}
\address[A]{J. R. Zubizarreta\\
Division of Decision,\\
\quad Risk and Operations\\
Department of Statistics\\
Columbia University\\
3022 Broadway\\
New York, New York 10027\\
USA\\
\printead{e1}} 
\address[B]{R. D. Paredes\\
Department of Industrial\\
\quad and Systems Engineering\\
Pontificia Universidad Cat\'{o}lica\\
Santiago de Chile\\
Chile\\
\printead{e2}}
\address[C]{P. R. Rosenbaum\\
Department of Statistics\\
The Wharton School\\
University of Pennsylvania\\
Philadelphia, Pennsylvania 19104-6340\\
USA\\
\printead{e3}}
\end{aug}
\thankstext{t1}{Supported in part by Grant 1110485 from Fondecyt and
Grant SBS-1260782 from the US National Science Foundation.}

\received{\smonth{7} \syear{2013}}
\revised{\smonth{12} \syear{2013}}

%
\begin{abstract}
Conventionally, the construction of a pair-matched sample
selects treated and control units and pairs them in a single step with
a view
to balancing observed covariates $\mathbf{x}$ and reducing the
heterogeneity or dispersion of treated-minus-control response differences,
$Y$. In contrast, the method of cardinality matching developed here
first selects the maximum number of units subject to covariate balance
constraints and, with a balanced sample for $\mathbf{x}$ in hand,
then separately pairs the units to minimize heterogeneity in $Y$.
Reduced heterogeneity of pair differences in responses $Y $ is
known to reduce sensitivity to unmeasured biases, so one might hope that
cardinality matching would succeed at both tasks, balancing $\mathbf{x}$,
stabilizing $Y$. We use cardinality matching in an
observational study of the effectiveness of for-profit and not-for-profit
private high schools in Chile---a controversial subject in Chile---focusing
on students who were in government run primary schools in 2004 but
then switched to private high schools. By pairing to minimize heterogeneity
in a cardinality match that has balanced covariates, a meaningful
reduction in
sensitivity to unmeasured biases is obtained.
\end{abstract}

%

\end{frontmatter}

\section{Introduction}

\subsection{Educational test scores and school profits}\label{ssIntroExample}
In Chile, as in the US, Britain, Canada and elsewhere, some secondary schools
are operated by the government and others are private enterprises that charge
parents a fee to educate their children. In Chile, some of the private
schools are not-for-profit enterprises, for instance, a~school operated
by a
church, and others are for-profit enterprises not different in concept
than a
restaurant or retail store. Whether schools should be allowed to
profit is
an intensely controversial issue in Chile. On the one hand,
supporters of
for-profit schools argue that they have incentives for efficiency and
innovation, and that this in turn results in better education. Opposing this
view, detractors say that, in reducing costs, for-profit schools tend
to also
reduce the quality of education and that one cannot allow a desire for
profits to take precedence over the quality of a child's education [see \citet{Ela09}
for further discussion]. In 2011,
in support of the latter view, and in part with the goal of ending for-profit
education in Chile, thousands of students rallied through the streets
demanding a change in the model of education and better opportunities.

Here, we compare the 2006 academic test performance of Chilean students who
entered for-profit private high schools and students who entered
not-for-profit private high schools. All of these students were in
government run primary/middle schools in Santiago in 2004 and subsequently
moved to private high schools. We have test scores at baseline in
2004 in
language (Spanish), mathematics, natural science and social science,
and we
have outcome test scores in 2006 in language and mathematics. In addition,
we have extensive data about parents and children in 2004, such as the
education of the parents, their income, the number of books at home and so
on, recorded in an observed covariate $\mathbf{x}$. An obvious
concern is
that even after adjusting for a high-dimensional observed covariate
$\mathbf{x}$, children in different types of schools may differ in
terms of
some other covariate $u$ that was not observed, and differences in $u$ may
bias the comparison.

The test scores come from the SIMCE, the Spanish acronym for ``System
of Measurement of Quality in Education.'' For the
same students, we use test scores for the 8th grade of primary school
in 2004
and the second year of high school in~2006. For the typical student, these
are test scores at ages 14 and 16. For-profit and not-for-profit are
determined by the official definitions of the Chilean IRS based on the
institutional identification number (RUT).

Do profits boost or depress test scores in similar students? Or are profits
irrelevant to test scores?

\subsection{Matching for covariate balance, pairing for
heterogeneity}\label{ssIntroPlan}

To be credible, the comparison must compare children in not-for-profit schools
(the treated group) to children similar at baseline in for-profit
schools (the
control group), and there are many ways the children may differ. It is
typically difficult to match closely for all coordinates of a high-dimensional
observed covariate $\mathbf{x}$, but it is often not difficult to create
matched treated and control groups with similar distributions of~$\mathbf{x}$.
For instance, if $\mathbf{x}$ consisted of 20 binary covariates, it would
distinguish $2^{20}$ or about a million categories of students, so it
would be
very difficult to match thousands of students exactly for all 20 covariates.
However, it is not difficult to balance $\mathbf{x}$ in treated and control
groups, for instance, by matching for an estimate of the one-dimensional
propensity score, that is, for an estimate of the conditional
probability of
treatment given the observed covariates [\citet{RosRub83}].
The
resulting matched pairs are heterogeneous in $\mathbf{x}$ but the
heterogeneity in $\mathbf{x}$ is unrelated to treatment and so tends to
balance out in the treated and control groups as whole groups. Randomized
treatment assignment also balances covariates without eliminating
heterogeneity in covariates, but of course randomization balances both
observed covariates $\mathbf{x}$ and unobserved covariate $u$, whereas
matching for the observed $\mathbf{x}$ cannot be expected to balance
$u$. It
is typically difficult to randomly assign students to schools, although
it has
happened in special situations.

If pairs matched for $\mathbf{x}$ have a not-for-profit-minus-for-profit
matched pair difference $Y$ in outcome test scores that is not centered at
zero, then the explanation may be an effect of
not-for-profit-versus-for-profit schools or it may instead reflect some
pretreatment difference in an unobserved covariate $u$. A sensitivity
analysis in an observational study asks: what would $u$ have to be like to
explain the observed behavior of $Y$ in the absence of a treatment effect?
In the first sensitivity analysis, \citet{Coretal59} found that to
explain away the observed association between heavy smoking and lung
cancer as
something other than an effect caused by smoking, the unobserved $u$ would
need to be a near perfect predictor of lung cancer and an order of magnitude
more common among smokers than nonsmokers. In Section~\ref
{ssSenAnalysis}, a
closely related though considerably more general method of sensitivity
analysis is reviewed.

It is known that the heterogeneity of $Y$, its dispersion around its center,
affects the degree of sensitivity to unmeasured biases [\citet{Ros05}]; see
Section~\ref{ssHeterogeneity} below. A typical effect of, say, $\tau$,
will be
more sensitive to an unobserved bias $u$ in treatment assignment if the $Y$'s
are widely dispersed about $\tau$ and less sensitive if the $Y$'s are tightly
packed around $\tau$, and this pattern will persist no matter how large the
sample size becomes. In this sense, reducing the heterogeneity or dispersion
of individual pair differences $Y$ is more important than increasing the
sample size, because an increase in sample size has little to do with
sensitivity to bias (or, more precisely, heterogeneity affects design
sensitivity but sample size does not). The heterogeneity of the $Y$'s is
partly determined by factors that the investigator cannot control, but often
the investigator has some control. To some extent, the heterogeneity
of $Y$
may be affected by the use of special populations, say, twins or
siblings who
happened to receive different treatments. To a limited extent, the
heterogeneity of the pair differences, $Y$, is affected by how the
pairing for
$\mathbf{x}$ is done. Our goal in the current paper is to reduce sensitivity
to unmeasured biases from $u$ by pairing in such a way that the heterogeneity
of $Y$ is reduced.

Conventionally, matching for $\mathbf{x}$ and pairing for $\mathbf{x}$ are
conceived as one task: treated and control groups are made similar as groups
in terms of $\mathbf{x}$ by pairing treated and control individuals with
similar $\mathbf{x}$'s. Using a new matching algorithm called
``cardinality matching'' in
Section~\ref{secCardinality}, we form matched treated and control
groups that are
of the largest proportional size possible (i.e., the maximum
cardinality) such
that the distributions of $\mathbf{x}$ are balanced in the groups as a whole.
The result is either the maximum number of pairs possible subject to
covariate balance constraints or the largest $L$-to-1 match using all treated
individuals, again subject to covariate balance constraints. In other words,
the marginal distributions of $\mathbf{x}$ in treated and control
groups are
constrained to be similar, and the maximum cardinality match is the largest
proportional match that makes them similar. The algorithm that
produces the
maximum cardinality match is indifferent as to who is paired with whom;
instead, it maximizes the size of a match that meets specified requirements
for balance on $\mathbf{x}$; see (\ref{eqOneConstraint}) below. This
is done
using integer programming. Then, with the groups determined and
fixed, pairs
or $L$-to-1 matched sets are formed using minimum distance pair
matching for a
robust Mahalanobis distance computed from a few key coordinates of
$\mathbf{x}$ with a view to reducing heterogeneity in the outcome
within pairs
or matched sets. An alternative approach is described in
Section~\ref{ssEnhancement}.

In the Chilean schools in Section~\ref{ssIntroExample}, pairs are
formed using test
scores in 2004, so treated and control groups are balanced for all of
$\mathbf{x}$ by maximum cardinality matching, yet individual pairs are also
paired very closely for 2004 test scores by optimal pair matching. In other
words, the treated and control groups have the same proportion of boys, the
same proportion of mothers who completed secondary school and so on, so the
treated and control groups look comparable as groups in terms of the measured
covariates. However, the pairing is concerned with test scores in middle
school, so a boy with good language scores and poor math scores may be paired
with a girl with similar test scores.

Unlike cardinality matching, typical matching algorithms find matched groups
that are balanced for $\mathbf{x}$ at the same time as they find pairs close
on $\mathbf{x}$. In doing this, typical algorithms do not usually
find the
largest matched sample that balances observed covariates; after all,
this is
not the criterion that they optimize. Additionally, typical
algorithms will
balance gender by trying to pair boys with boys, even if gender is not a
strong predictor of test performance in high school. If one is going to
break up the initial pairing and pair the same individuals a second time
(henceforth, if one is going to ``re-pair''),
then effort spent making the initial pairing close on $\mathbf{x}$ is effort
wasted; after all, the initial pairing is not used. Cardinality
matching is
most attractive when a convincing comparison must balance many covariates,
even though it is known that a small subset of the covariates is key for
predicting the outcome. Cardinality matching is least attractive when there
is no reason to think that some covariates or covariate summaries are much
more important for prediction than others.

The key covariates for revised pairing are known before the study
begins in
many contexts. This is true, for example, of the baseline 2004 test scores
in the Chilean schools in Section~\ref{ssIntroExample}, and it is also
true of
clinical stage, grade and histology in some clinical cancer studies. In
other contexts, there are widely used, extensively validated summary scores
that could be used for the revised pairing, such as the APACHE score in
clinical medicine [\citet{Knaetal85}] or the Charleson Index in health
services research [\citet{DeyCheCio92}]. Obviously, one can match for
both such
a summary score and a few key covariates using some form of the Mahalanobis
distance. \citet{Rub79} found that covariance adjustment of matched pair
differences is a particularly robust technique, being little affected by
misspecification of the regression model, and his approach using all of
$\mathbf{x}$ can additionally provide some insurance against an
omission when
identifying the key covariates for revised pairing. Sensitivity analysis
after covariance adjustment of matched pairs is illustrated in
\citet{Ros07}.

\citet{Bai11} proposed re-pairing any initial pair-matched sample by, first,
using the unused, unmatched controls to estimate Hansen's (\citeyear{Han08}) prognostic
score, and, second, revising the initial pairing to be close on the estimated
prognostic score, so that, after revision, pairs have similar predicted
responses under control. Baiocchi's revised match retains whatever balancing
properties for $\mathbf{x}$ that the initial match may have had,
because it
uses the same treated and control groups, yet the new pairs are now
close in
terms of a prognostic score whose estimated weights came from data independent
of the paired data that will be the basis for the study's conclusion. A
limited version of Baiocchi's method would simply use the unused, unmatched
controls to identify the most important covariates for predicting the outcome
and then re-pair using those covariates directly. Baiocchi's method concerns
the second step, the revision of a balanced match, and it is a natural
complement to cardinality matching that concerns the first step, namely,
finding the largest balanced matched sample ignoring who is matched to whom.
The key variables for revised pairing are known a priori in some contexts,
but when this is not the case, Baiocchi's method is a clever and useful
strategy for revising the pairing of a balanced matched sample.

Reducing the dispersion or heterogeneity of pair differences $Y$ reduces
sensitivity to unmeasured biases, but increasing the sample size does not.
Is matching each treated subject to $L>1$ controls analogous to reducing
heterogeneity or to increasing the sample size? Matching with more
than one
control often reduces sensitivity to unmeasured biases [\citet{Ros13}].
Stated informally, this occurs when an unmeasured covariate $u$
cannot both
closely predict the pattern of outcomes among $L+1$ individuals in an $L
$-to-1 matched set and also closely predict which one of $L+1$ individuals
will receive the treatment. When possible, cardinality matching will
automatically construct $L$-to-1 matched sets with the largest $L$ if
this is
consistent with balancing $\mathbf{x}$, and otherwise it will find the largest
1-to-1 pair matching that balances $\mathbf{x}$.

For recent surveys of multivariate matching, see \citet{Stu10} and \citet{Luetal11}.

\subsection{Outline and key ideas}
The remainder of the paper discusses and illustrates the following
three topics.

\begin{description}
\item \textit{A new method}: The visible heterogeneity of responses within matched
pairs affects the sensitivity of conclusions to unmeasured biases [\citet{Ros05}]. A new matching algorithm, cardinality matching, balances many
covariates but pairs for just a few covariates that reduce the heterogeneity
of matched pair differences in outcomes, thereby reducing sensitivity to
unmeasured biases. Cardinality matching finds the largest match that meets
the user's specifications for covariate balance, also addressing the
possibility of covariate distributions exhibiting limited overlap.

\item \textit{Recent developments}: A poor choice of test statistic can lead
to a
mistaken view that an observational study is sensitive to small biases
when it
is not. We illustrate an adaptive choice of test statistic in sensitivity
analysis [\citet{Ros12N1}].

\item \textit{A case study}: The case study of for-profit schools in Chile
illustrates cardinality matching and the switch from a conventional
match and
analysis to an alternative guided by statistical theory produces a substantial
reduction in reported sensitivity to unmeasured biases.
\end{description}

Section~\ref{secCardinality} describes the new matching algorithm and
Section~\ref{secSensitivity} is a review of recent developments in sensitivity
analysis. Technical details may be avoided by focusing on the case
study in
Sections~\ref{ssStep1CardinalityMatch}, \ref{ssStep2optimalpairing},
\ref{ssPrelimAnalysis} and \ref{secSenInExample}.

\subsection{Aspects of the Chilean data}\label{ssData}

We compare test scores of students in Santiago who moved from a public primary
school in 2004 to either a private for-profit or a private non-for-profit
secondary school in 2006. The data are from the Education Quality Measurement
System (SIMCE) which contains results from a standardized test given by the
Ministry of Education to all the students in Chile in a given year. Unlike
standardized educational tests in the US, the SIMCE tests every student in
Chile and in this sense resembles a census rather than a sample or an
administrative data set. After applying basic data exclusion criteria
[namely, excluding from the analysis those students (i) who were not in
Santiago, (ii) who did not move from a public primary school in 2004 to either
a private for-profit or a non-for-profit secondary school in 2006,
(iii) whose
reported gender changed between years, or (iv) who had missing values
in one
of the baseline or outcome test scores], before matching we obtained
data from
students in 483 public primary schools in 2004. After matching, our matching
algorithm selected students from 446 of these 483 public primary
schools. The
sample of matched students had students from 453 private secondary
schools in
2006 (170 for-profit and 283 non-for-profit). Before matching there
were 573
private secondary schools, 170~for-profit and 403 non-for-profit.

\section{Cardinality matching followed by minimum distance
pairing}\label{secCardinality}

\subsection{Cardinality matching: The largest matched sample that
balances covariates}\label{ssCardinalityMethod}

Cardinality matching finds the largest match that balances observed
covariates. Balancing observed covariates is expressed abstractly by $K$
linear inequalities in functions of the observed covariates. Just as
it is
convenient to describe linear regression abstractly, and then later observe
that the abstract definition permits interactions, polynomials, some
types of
splines, nominal predictors, etc., so too it is convenient to describe
covariate balance abstractly, and then observe that various ways of
making the
abstract statement tangible may be used to achieve a variety of desirable
effects. For instance, the $K$ linear inequalities can balance proportions,
means, variances, covariances, and a grid of quantiles of a marginal
distribution,
among many other effects.

There are initially treated units $\mathcal{T}= \{ \rho_{1},\ldots,\rho
_{T} \} $ and controls $\mathcal{C}= \{ \kappa_{1},\ldots,\allowbreak\kappa_{C} \}
$. Treated unit $\rho_{t}$ has observed
covariate $\mathbf{x}_{\rho t}$, $t=1,\ldots,T$, and control $\kappa
_{c}$ has
observed covariate $\mathbf{x}_{\kappa c}$, $c=1,\ldots,C$. Let $a_{tc}=1$
if $\rho_{t}$ is initially matched to $\kappa_{c}$, with $a_{tc}=0$ otherwise.
Each matched treated unit is to have the same number, $L\geq1$, of matched
controls, where the algorithm will make $L$ as large as possible
subject to
the requirement that the covariates be balanced in treated and control groups.
More precisely, it will either find the largest match using all $T$ treated
units each matched to $L$ distinct controls or it will find the 1-to-1
matching that uses the maximum number of treated units. A covariate balance
constraint $\mathbb{B}_{k}$ is a linear inequality constraint
%
%
\begin{equation}
\mathbb{B}_{k}\dvtx-b_{k}\sum_{t=1}^{T}
\sum_{c=1}^{C}a_{tc}\leq\sum
_{t=1}^{T}%
\sum
_{c=1}^{C}a_{tc} v_{ktc}\leq
b_{k}\sum_{t=1}^{T}\sum
_{c=1}^{C}%
a_{tc},
\label{eqOneConstraint}%
\end{equation}
where\vspace*{1.5pt} $v_{ktc}$ is the $k$th of $K$ functions of observed covariates and
$b_{k}\geq0$ is a given constant. Specifically,\vspace*{1pt} $\mathbb{B}_{k}$ says the
mean $ ( \sum_{t=1}^{T}\sum_{c=1}^{C}a_{tc} v_{ktc} ) /\allowbreak (
\sum_{t=1}^{T}\sum_{c=1}^{C}a_{tc} ) $ of~$v_{ktc}$ over matched units
($a_{tc}=1$) is in the interval $ [ -b_{k}, b_{k} ] $, and taking
$b_{k}=0$ says the mean of $v_{ktc}$ over matched units ($a_{tc}=1$) is zero.

\begin{figure}[t]
\includegraphics{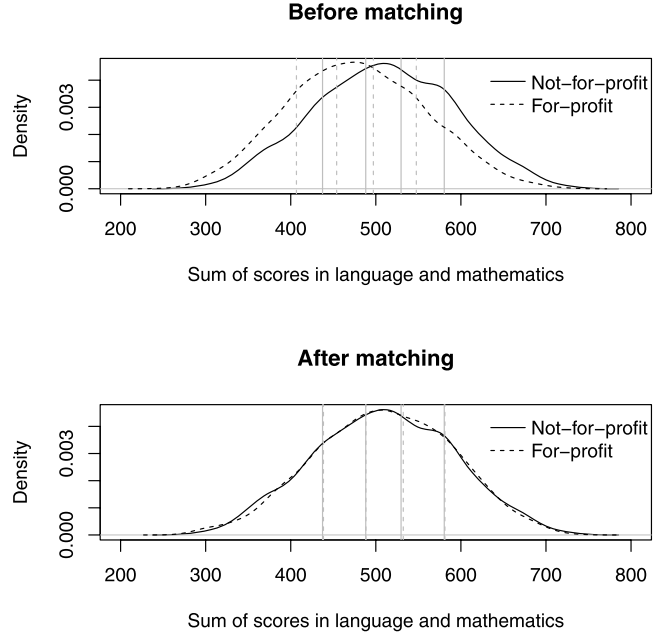}
\caption{Total of language and mathematics scores at baseline in
2004, before and after cardinality matching. Vertical lines indicate
quintiles.}\label{fig1}
\end{figure}

Many useful balance constraints have the form (\ref{eqOneConstraint}) with
$v_{ktc}=f ( \mathbf{x}_{\rho t} ) -f ( \mathbf{x}_{\kappa
c} ) $ for some function $f ( \cdot) $. If $f (
\cdot) $ is a binary indicator of whether $\mathbf{x}$ satisfies some
condition, then (\ref{eqOneConstraint}) with $b_{k}=0$ forces the matched
sample to have the same number of treated subjects satisfying this condition
as controls satisfying this condition, without constraining who is
matched to
whom. The covariates gender, school type, categories of household income,
and categories of mother's and father's education were exactly balanced in
this way, a constraint known as ``fine
balance'' [\citet{Zubetal11}]. Fine balance for
gender means that the proportion of boys is the same in the matched treated
and control groups, but boys may be paired with girls. When several
covariates are finely balanced, the mean of every linear combination of these
covariates is also exactly balanced. A binary indicator $f (
\cdot) $ with $b_{k}=0.01$, say, will limit the imbalance to at
most a
count of 1\%, a condition known as ``near fine
balance'' [\citet{Yanetal12}]. The categories of
``number of books at home'' were nearly
balanced in this way. In parallel, $f ( \cdot) $ with $b_{k}=0$
may be used to balance the joint distributions of two or more nominal
covariates, say, the gender of the student and the years of education
of the
mother. If $f ( \cdot) $ simply picks out one coordinate of
$\mathbf{x}$, then a pair of constraints of the form (\ref{eqOneConstraint})
forces the matched sample to have means in the treated and control
groups that
differ by at most $b_{k}$, say, that the mean test scores in natural
science in
2004 are close. The student's own four test scores in 2004 and the four
average test scores in the student's 2004 school were balanced on
average in
this way. If instead $f ( \cdot) $ calculates the square of one
coordinate or the cross-product of two coordinates, then a sequence of
constraints of the form (\ref{eqOneConstraint}) can balance higher
moments of
the covariates. A binary indicator $f ( \cdot) $ may be
used to
ensure that the same number or a similar number of treated subjects and
controls have a value of one covariate below a particular number, and a
sequence of such binary indicators may be used to force agreement
between two
empirical distribution functions at the grid of values. In Figure~\ref{fig1}, the
entire distribution of the sum of math and language scores in 2004 was
balanced in this way. In an analogous way, constraints of the form
(\ref{eqOneConstraint}) may be used to ensure that an estimated propensity
score has a similar distribution in treated and control matched samples.
Also, rather than eliminate subjects with missing covariates, one can force
treated and control matched groups to exhibit similar patterns of missing
covariates, say, 5\% of a particular covariate being missing in both groups.
For detailed discussion of the variety of statistical properties that
may be
induced through balance constraints of different types, see
\citet{Zub12}.

The user of cardinality matching specifies $K$ constraints of the form
(\ref{eqOneConstraint}). The goal is to find the largest $L$-to-1
match that
satisfies the $K$ balance constraints, the largest match that balances
all of
the observed covariates. The result may be, say, a 3-to-1 match of all
treated units, or it may be a 1-to-1 pair match discarding the smallest
possible fraction of the treated units. In any case, the algorithm
finds the
largest $L$-to-1 match that exists subject to the $K$ constraints that define
covariate balance. A cardinality matching is then the solution to the
following several optimization problems. First, find $\mathbf{a}= (
a_{11},a_{12},\ldots,a_{TC} ) $ as the solution to
%
%
\begin{eqnarray}\label{eqCardMatch}
&& \max\sum_{t=1}^{T}\sum_{c=1}^{C}a_{tc}\nonumber
\\
&& \mbox{subject to } a_{tc}  \in \{ 0,1 \},\qquad t=1,\ldots,T, c=1,\ldots,C,\nonumber
\\
&&\hspace*{45pt} \sum_{t=1}^{T}a_{tc} \leq 1\qquad\mbox{for }c=1,\ldots,C,
\\
&&\hspace*{45pt} \sum_{c=1}^{C}a_{tc}  \leq 1 \qquad\mbox{for }t=1,\ldots,T,\nonumber
\\
&&\hspace*{45pt} \mathbb{B}_{k},\qquad k  = 1,\ldots,K.\nonumber
\end{eqnarray}
In words, (\ref{eqCardMatch}) is the largest pair-matched sample that meets
the user's $K$ balance constraints $\mathbb{B}_{k}$, $k=1,\ldots,K$ in
(\ref{eqOneConstraint}). Specifically, $\sum_{t=1}^{T}\sum_{c=1}^{C}a_{tc}$
is the number of subjects in the treated and control groups, $\sum_{t=1}
^{T}a_{tc}\leq1$ says that control $c$ is used at most once, and $\sum
_{c=1}^{C}a_{tc}\leq1$ says treated unit $t$ is used at most once.

Having solved (\ref{eqCardMatch}), there are two cases to consider. In case
1, the solution to~(\ref{eqCardMatch}) has $T=\sum_{t=1}^{T}\sum_{c=1}%
^{C}a_{tc}$, so that a pair match satisfying the balance constrains~$\mathbb{B}_{k}$,
$k=1,\ldots,K$ constraints has been found that uses
all $T$
treated units. In this first case, the problem is solved again with the
third constraint, $\sum_{c=1}^{C}a_{tc}\leq1$ for $t=1,\ldots,T$,
replaced by
$\sum_{c=1}^{C}a_{tc}=L$ with $L=2$ for $t=1,\ldots,T$. If this second
solution has $LT=2T=\sum_{t=1}^{T}\sum_{c=1}^{C}a_{tc}$, then a $2$-to-$1$
match satisfying the balance constrains $\mathbb{B}_{k}$, $k=1,\ldots,K$
constraints has been found, and the problem is solved again with $L$ replaced
by $L+1$. For some $L$, $L=2$, $3, \ldots,$ the problem is infeasible,
meaning that a match of $L$-to-$1$ cannot satisfy the balance constraints~$\mathbb{B}_{k}$, $k=1,\ldots,K$. In this first case, the optimal
cardinality match is the feasible solution with the largest $L$
satisfying the
balance constrains $\mathbb{B}_{k}$, $k=1,\ldots,K$. In case 2, if the
solution to (\ref{eqCardMatch}) has $T>\sum_{t=1}^{T}\sum_{c=1}^{C}a_{tc}$,
then even a $1$-to-$1$ pair match that uses all $T$ treated units will violate
the balance constrains~$\mathbb{B}_{k}$, $k=1,\ldots,K$ constraints,
and the
algorithm has found the largest 1-to-1 pair matching that does satisfy the
balance constraints. [In the abstract, one should solve (\ref{eqCardMatch})
and the adjusted match for every integer $2\leq L\leq C/T$, but in realistic
practice it is very unlikely that a feasible solution exists for
$L^{\prime}>L$ if there is no feasible solution for $L$.]

Cardinality matching differs from optimal matching [\citet{Ros87}]
in that
its objective function $\sum_{t=1}^{T}\sum_{c=1}^{C}a_{tc}$ in
(\ref{eqCardMatch}) is simply the size of a matched sample that satisfies
balance constraints (\ref{eqOneConstraint}), whereas optimal matching
has as
its objective $\sum_{t=1}^{T}\sum_{c=1}^{C}a_{tc} \eta_{tc}$, where
$\eta
_{tc}$ is a measure of the distance between $\mathbf{x}_{\rho t}$ and
$\mathbf{x}_{\kappa c}$, typically a Mahalanobis distance with a
caliper on
the propensity score implemented using a penalty function [e.g., \citet{Ros10N1}, Section~8]. In cardinality matching, the balance constraints,
$\mathbb{B}_{k}$, $k=1,\ldots,K$, refer only to the marginal
distributions of
$\mathbf{x}$ in matched samples, so the pairing of treated and control
subjects is arbitrary, in the sense that none of the quantities that define
the optimization problem (\ref{eqCardMatch}) are affected by who is paired
with whom. The approach we take here is to solve (\ref{eqCardMatch}) using
only constraints on distributions of $\mathbf{x}$ in treated and control
groups, thereby obtaining the largest balanced matched samples; then,
with the
matched sample fixed, we re-pair units within the sample to minimize a
distance, $\sum\sum a_{tc} \eta_{tc}$, over the fixed matched sample.
The
advantage of the two-step approach is that (\ref{eqCardMatch}) will yield
treated and control groups that look comparable in terms of observed
covariates $\mathbf{x}$; then, pairing to minimize $\sum\sum a_{tc}
\eta
_{tc}$ will focus on reducing heterogeneity in $Y$, where reducing
heterogeneity in $Y$ can reduce sensitivity to unmeasured biases.

Traditionally, in experimental design, randomization balanced
covariates and
prevented bias, while blocking or pairing for covariates increased efficiency;
see, for instance, \citet{Cox58}. In a somewhat parallel way, cardinality
matching balances observed covariates while pairing following cardinality
matching reduces heterogeneity. The key distinction is randomization
addresses biases from unmeasured covariates where cardinality matching does
not, and a reduction in heterogeneity affects sensitivity to biases from
unmeasured covariates, these biases being absent in a randomized experiment.

\subsection{Step 1: Cardinality matching in Santiago using covariates in 2004}\label{ssStep1CardinalityMatch}

Solving (\ref{eqCardMatch}) yielded a maximum of $\sum_{t=1}^{T}\sum
_{c=1}%
^{C}a_{tc}=1907$, meaning 1907 pairs of a treated and control subject
satisfying the balance constraints. Because $\sum_{t=1}^{T}\sum_{c=1}%
^{C}a_{tc}=1907=T$, all $T=1907$ of the treated students were matched,
and the
method in Section~\ref{ssCardinalityMethod} then tried to construct a
2-to-1 match
subject to the same balance constraints. However, no 2-to-1 match satisfies
the balance constraints, that is, the second step of the optimization problem
is infeasible. The largest $L$-to-1 match that balances the
covariates is a
1-to-1 match that uses all the treated students.

The for-profit and not-for-profit matched groups had exactly the same number
of men (855 men in both groups) and women (1052 women in both groups), exactly
the same number of people from each of four zones of Santiago, exactly the
same number from each of seven categories of household income, exactly the
same number with each of five categories of mother's education, and
exactly the
same number with each of five categories of father's eduction. For income,
mother's and father's education, one of the categories was
``missing,'' and ``missing''
was balanced. Most of these covariates were ``finely
balanced'' in the sense that the distributions were exactly
the same in for-profit and not-for-profit groups, but the two
individuals in a
pair may differ with respect to the covariate.

Other covariates were constrained to have distributions that were very similar
but not identical in means or proportions. For instance, the mean of the
baseline language${}+{}$mathematics score was 509.05 in the for-profit group and
509.16 in the not-for-profit group. The baseline test scores in language,
mathematics, natural science and social science were similarly mean-balanced.
The average test scores in a student's school give some indication of the
student's peers at school, and each student has school averages in language
(Spanish), mathematics, natural and social science. These school average
scores were similarly mean-balanced. The number of books in a
student's home
was represented by six categories, from none to more than 200, and the
proportions were closely balanced. For all of these covariates, the
for-profit-minus-not-for-profit difference in covariate means or proportions
was at most 6 one hundredths of the standard deviation of the variable before
matching. An online supplement describes the covariate balance in detail
[\citet{autokey51}].

Cardinality matching ended up using all 1907 treated students in 1907 matched
pairs, but in some other problem it might use a subset of treated
students in
its effort to satisfy the balance constraints $\mathbb{B}_{k}$,
$k=1,\ldots,K$. That is, if the treated group and the potential
controls have a limited
region of overlap on observed covariates, cardinality matching might
produce a
subset match confined to the region of overlap, thereby ensuring covariate
balance. For other methods of subset matching, see \citet{Cruetal09},
\citet{TraSma}, \citet{Ros12N2} and \citet{HilSu13}.

\subsection{Step 2: Optimal pairing of a given match using covariates
in 2004}\label{ssStep2optimalpairing}

To illustrate the advantages of separating balancing of covariates and pairing
of individual students, the one match in Section~\ref
{ssStep1CardinalityMatch} is
paired in two different ways to form two sets of 1907 pairs. To emphasize,
the same $2\times1907$ students are paired, but who is paired with whom is
different in the two pairings. Because the treated and control groups
do not
change, covariate balance is identical in both pairings, because covariate
balance ignores who is paired with whom. The first pairing uses a robust
Mahalanobis distance [\citeauthor{Ros10N1} (\citeyear{Ros10N1}), Section~8.3] based on all of
the covariates
used in (\ref{eqCardMatch}), so it views test scores, parents' education,
books at home, etc., as equally important. The second pairing uses the
robust Mahalanobis distance but computed just from the four baseline test
scores. In both matches, the total of the 1907 covariate distances within
pairs is minimized using the optimal assignment algorithm, as might be done,
for example, using the \texttt{pairmatch} function of Hansen's (\citeyear{Han07})
\texttt{optmatch} package in \texttt{R}. One pairing yields pairs
that are
somewhat close on all covariates; the other pairing yields pairs that
are very
close on test scores, being content to balance the other covariates.
Although one would not want to compare groups of students whose
parents had
very different levels of education or very different numbers of books
at home,
it is generally the case that test scores best predict related test scores.

\begin{figure}

\includegraphics{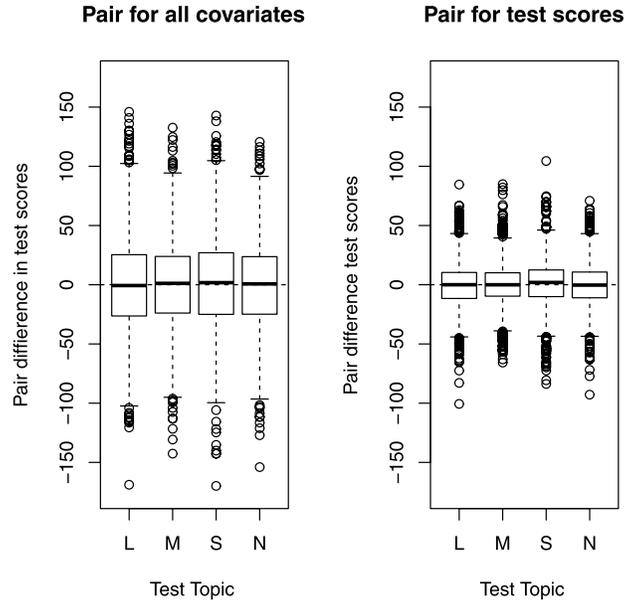}

\caption{Comparison of two ways of pairing the same students.
Treated-minus-control pair differences in test scores for 1907 pairs at
pretreatment baseline in 2004 in four subject areas, $\mathrm{L}={}$language,
$\mathrm{M}={}$mathematics, $\mathrm{S}=$ social science, $\mathrm{N}={}$natural
science. The same 1907 treated students and 1907 control students are in
both pairings, but the pairing on the right emphasized pairing for baseline
test scores, whereas the pairing on the left gave equal emphasis to all
baseline covariates.}\label{fig2}
\end{figure}

Figure~\ref{fig2} depicts the pair differences in the four test scores in 2004, when
all $1907\times2=3814$ were attending government run primary/middle schools.
On the left in Figure~\ref{fig2}, the pairing used all covariates, whereas on the
right the pairing focused on test scores. On both the left and the right,
the distribution of treated-minus-control differences is centered at zero,
because the matching in Section~\ref{ssStep1CardinalityMatch} balanced the
distributions of test scores. As expected, when the pairing focused
on test
scores, the baseline difference in test scores was closer to zero, that
is, on
the right in Figure~\ref{fig2}, the boxplots are more compact about zero. Of course,
other covariates are further apart within pairs when pairing emphasizes test
scores, but the distributions of these other covariates are equally balanced
for both pairings in Figure~\ref{fig2}.

\subsection{Comparison with cem: Coarsened exact matching}\label{ssCompareCEM}

Coarsened exact matching (or \texttt{cem} in \texttt{R}) is a popular, recent
proposal for matching that finds pairs close on $\mathbf{x}$; see \citet{IacKinPor}.
At the suggestion of a referee, we compare cardinality pair matching
to pair matching using \texttt{cem}. Essentially, it rounds or
coarsens each
coordinate of $\mathbf{x}$, makes strata that are homogeneous in all of the
coarsened coordinates, and eliminates all strata that do not contain at least
one treated subject and one control. To the extent that \texttt{cem}
balances covariates, it does this by making the pairs individually
close on
each coordinate of $\mathbf{x} $. One expects the performance of
\texttt{cem} to vary with the dimensionality of $\mathbf{x}$, among other
considerations, and the dimensionality of $\mathbf{x}$ strongly
affected the
performance of \texttt{cem} in the current example.

Using the default settings in \texttt{R} and matching for all of the
categorical and continuous covariates balanced by cardinality matching,
\texttt{cem} produced 3 matched pairs, as opposed to 1907 pairs by cardinality
matching. That is, there were only 3 treated students who fell in the same
coarsened exact stratum as a control. The default for \texttt{cem} is 12
categories for a continuous covariate, however, if this is reduced to 4
categories, then \texttt{cem} produced 21 matched pairs.

When coarsened exact matching is used with fewer covariates it produces fewer,
denser strata and many more pairs. We estimated a propensity score
using all
of the covariates to predict treatment assignment in a logit model. When
used with just two covariates, the total of the four baseline test
scores and
the estimated propensity score, \texttt{cem} produced 1856 of a
possible 1907
pairs. In theory, matching for a well-estimated propensity score should
balance all the observed covariates in the score in a stochastic sense, much
as coin flips tend to balance covariates in randomized experiments. Matching
for the propensity score did a tolerable job of stochastically
balancing many
covariates, but, unlike the perfect balance obtained by cardinality matching,
there were some nominal covariates that differed significantly, as is expected
with many covariates even in a randomized experiment, for instance, mother's
education differed significantly in for-profit and not-for-profit groups.

How did cardinality matching compare with the two-covariate \texttt{cem}
match? Presumably, either could be used in practice. However, the
cardinality match produced better covariate balance and more matched pairs.

\subsection{An enhancement of cardinality matching: The closest largest
balanced match}\label{ssEnhancement}

\begin{figure}[b]

\includegraphics{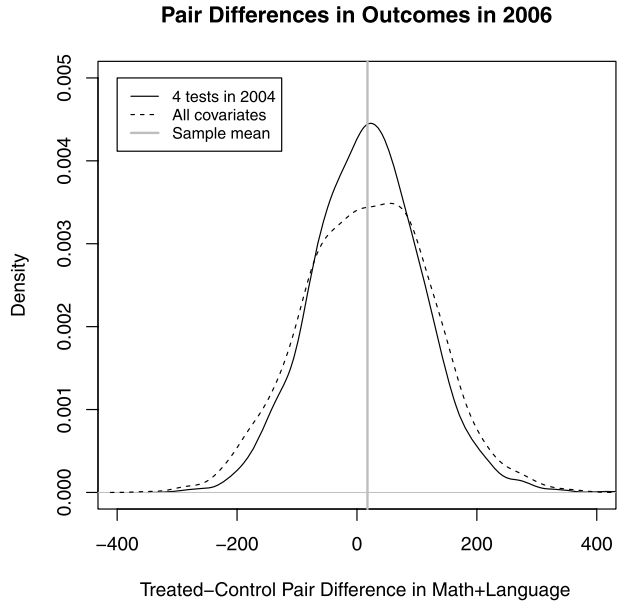}

\caption{Density estimate of 1907 matched pair differences in 2006
outcomes pairing either for the four 2004 baseline test scores or for all
covariates. Because the same $2 \times 1907 = 3814$ students appear in
both paired comparisons, the mean difference is the same, 17.5 points. The
dispersion of the pair differences is smaller when pairing for the four
2004 test scores: standard deviation of 90.9 versus 105.5, MAD of 60.2
versus 72.6.}\label{fig3}
\end{figure}

In principle, the method in Section~\ref{ssCardinalityMethod} may be
improved at
the price of some additional computation. In the Chilean schools example,
the computational effort increased without benefit, but, in a formal
sense, the
enhanced match is as large as the match in Section~\ref{ssStep1CardinalityMatch}
and satisfies the same $K$ balance constraints~(\ref{eqOneConstraint}), but
might possibly be closer in the second step in Section~\ref
{ssStep2optimalpairing}.
In principle, there may be more than one, perhaps many, $L$-to-1 balanced
matched samples of maximum cardinality, that is, many solutions $\mathbf{a}$
to (\ref{eqCardMatch}) that satisfy the balance constraints $\mathbb{B}_{k}$,
$k=1,\ldots,K$ with the same $L$ and $\sum_{t=1}^{T}\sum_{c=1}^{C}a_{tc}$.
These several matches, when they exist, will have selected the same number
of controls but different individual controls, while satisfying the same
balance constraints. When this is true, it seems natural to prefer from
among these solutions $\mathbf{a}$ one that minimizes the distance
$\sum\sum a_{tc} \eta_{tc}$ used to control heterogeneity. This may be
done in a
straightforward way using a relatively standard device. First, one solves
the problem in Section~\ref{ssCardinalityMethod}, thereby determining
the size,
$n=\sum_{t=1}^{T}\sum_{c=1}^{C}a_{tc}$ and $L=\max( 1,n/T )
$, of
the largest $L$-to-1 match that satisfies the balance constraints
$\mathbb{B}_{k}$, $k=1,\ldots,K$ in the sense of Section~\ref
{ssCardinalityMethod}.
Then, this match is discarded---it serves simply to determine the
size of the largest match that satisfies the balance constraint $\mathbb
{B}_{k}$,
$k=1,\ldots,K$. One then solves the optimization problem that minimizes
$\sum\sum a_{tc} \eta_{tc}$ subject to the balance\vspace*{1pt} constraints $\mathbb
{B}%
_{k}$, $k=1,\ldots,K$ together with the constraint that it be an
$L$-to-1 size
$n=\sum_{t=1}^{T}\sum_{c=1}^{C}a_{tc}$ match. This problem is known
to be
feasible because the method in Section~\ref{ssCardinalityMethod} has
already found
one feasible solution. The solution to the second problem is not only the
largest $L$-to-1 matched sample that satisfies the balance constraints but
also, among all such matched samples, it is the closest, minimizing
$\sum\sum
a_{tc} \eta_{tc}$. We tried this method in the example. Of course, it
again produced $n=1907$ pairs satisfying $\mathbb{B}_{k}$, $k=1,\ldots,K$,
thereby producing virtually the same covariate balance; moreover, it reduced
$\sum\sum a_{tc} \eta_{tc}$ very slightly with virtually the same substantive
conclusions. We did not report this alternative match because it did not
permit the comparison of two matches of the same individuals in Figure~\ref{fig3}.

A practical disadvantage of the enhanced approach is that it requires the
distances $\eta_{tc}$ that are used to reduce heterogeneity to be determined
before the final controls are selected because the enhanced approach uses
those distances both in selecting and pairing controls. In
particular, this
precludes using Baiocchi's (\citeyear{Bai11}) promising method, described in
Section~\ref{ssIntroPlan}, in which the unmatched controls are used to estimate
Hansen's (\citeyear{Han08}) prognostic score which then is used to define~$\eta_{tc}$.

\subsection{Preliminary examination of results in 2006}\label{ssPrelimAnalysis}

In 2006, there are language and mathematics scores for students in a
not-for-profit (treated) or a for-profit (control) high school, where these
students were in a government-run primary school in 2004. Figure~\ref{fig3} depicts
the treated-minus-control pair differences $Y$ in total test scores in 2006,
the sum of language and mathematics. Specifically, Figure~\ref{fig3} is a density
estimate of the $Y$'s from the two pairings (obtained using \texttt{density}
in \texttt{R} with default settings). The mean pair difference in
2006 test
scores is, of course, the same for the two pairings, namely, 17.5 points,
because the mean difference equals the difference of the means, and the two
pairings have the same students paired differently. In contrast, the second
pairing that emphasized pretreatment 2004 test scores has yielded less
dispersion in 2006 difference in posttreatment test scores $Y$. This is
visible in Figure~\ref{fig3} in the density estimates of $Y$ in the two pairings.
Also, in the first pairing, the standard deviation and median absolute
deviation from the median (MAD) of $Y$ were 105.5 and 72.6 points,
respectively, whereas in the second pairing that emphasized pairing for 2004
test scores, the standard deviation and MAD of $Y$ were 90.9 and 60.2.
In
terms of the appearance of the density estimate in Figure~\ref{fig3}, in terms
of the
standard deviation and in terms of the MAD, the treated-minus-control
difference $Y$ in outcomes is more stable, less dispersed, when the pairing
emphasizes the pretreatment 2004 test scores. A~reduction in
dispersion of
$Y$ is expected to translate into reduced sensitivity to unmeasured biases
[\citet{Ros05}], a topic examined in detail in Section~\ref{secSensitivity}.

The pattern in Figure~\ref{fig3} is not surprising. Before pairing, ignoring
treatment, among the 3814 students in the cardinality match, the Spearman
correlation between income and total test score (mathematics${}+{}$language)
in 2006
was 0.195, whereas the correlations with pretreatment 2004 test scores in
social science and natural science were 0.632 and 0.604, respectively, while
the correlation with total test score (mathematics${}+{}$language) in 2004
was 0.727.

Is a difference of 17.5 points a consequential difference? It is 0.16 times
the \textit{population} standard deviation of the total of math and language
scores. An observational study by \citet{Bel09} of lengthening the school
day in Chile from half a day to a full day estimated an effect on language
scores of 0.06 times the standard deviation. Various studies in the
US of
the effectiveness of urban charter schools versus public schools have produced
estimates around 0.20 times the standard deviation; see \citet{AngPat}, page~1.

In short, the not-for-profit schools have higher test performance for students
who appeared similar in 2004 in terms of observed covariates $\mathbf{x}$.
The mean difference in outcomes $Y$ is 17.5 points in both pairings,
but the
$Y$'s are less heterogeneous, less dispersed, more stable in the
pairing that
focused on pretreatment test scores. Did reduced heterogeneity in $Y$ have
any effect on sensitivity to unmeasured biases?\vadjust{\goodbreak}

\section{Review of sensitivity analysis}\label{secSensitivity}

\subsection{Notation for randomized experiments}

There are $I$ matched pairs, $i=1,\ldots,I$, with two subjects in each pair,
$j=1,2$, one treated with $Z_{ij}=1$, the other control with
$Z_{ij}=0$. In
Section~\ref{ssIntroExample}, there are $I=1907$ pairs of two students,
one who
moved to a not-for-profit private school, $Z_{ij}=1$, the other who
moved to a
for-profit private school, $Z_{ij}=0$. Matched treated and control grouped
balanced observed covariates $\mathbf{x}_{ij}$ but may differ systematically
in terms of an unobserved covariate $u_{ij}$. Let $\mathcal{Z}$ be
the set
of possible values of $\mathbf{Z}= ( Z_{11},\ldots,Z_{I2} ) ^{T}$,
so $\mathbf{z}\in\mathcal{Z}$ if and only if $z_{ij}=0$ or $z_{ij}=1$ with
$z_{i1}+z_{i2}=1$ for all $i$. Conditioning on $\mathbf{Z}\in\mathcal
{Z}$ is
abbreviated as conditioning on $\mathcal{Z}$. Write $\llvert
\mathcal{S}\rrvert$ for the number of elements in a finite set, so
$\llvert\mathcal{Z}\rrvert=2^{I}$.

As in \citet{Neyman1923} and \citet{Rub}, each subject has two potential
responses, $r_{Tij}$ if treated with $Z_{ij}=1$, $r_{Cij}$ if control with
$Z_{ij}=0$, so response $R_{ij}=Z_{ij} r_{Tij}+ ( 1-Z_{ij} )
r_{Cij}$ is observed from $ij$ and the effect of the treatment on $ij$,
namely, $r_{Tij}-r_{Cij}$, is not observed. In Section~\ref{secCardinality},
$r_{Tij}$ is the total 2006 test score student $ij$ would exhibit in a
not-for-profit school, $r_{Cij}$ is the total 2006 test score this same
student $ij$ would exhibit in a for-profit school, $r_{Tij}-r_{Cij}$ is the
effect of not-for-profit-versus-for-profit on this one student, and $R_{ij}$
is the observed 2006 test score of student $ij$ in the type of school $Z_{ij}
$ that $ij$ actually attended. Write $\mathcal{F=} \{ (
r_{Tij}, r_{Cij}, \mathbf{x}_{ij}, u_{ij} ), i=1,\ldots,I, j=1,2 \} $.
Fisher's (\citeyear{Fis35}) sharp null hypothesis $H_{0}$
of no
treatment effect asserts $H_{0}\dvtx r_{Tij}=r_{Cij},\ \forall i,j$. Write
$\mathbf{R}= ( R_{11},\ldots,R_{I2} ) ^{T}$ and $\mathbf{r}%
_{C}= ( r_{C11},\ldots,r_{CI2} ) ^{T}$, so $\mathbf{R}%
=\mathbf{r}_{C}$ if $H_{0}$ is true.

In a randomized paired experiment, treatments are assigned
independently by
the flip of a fair coin, so $\Pr( \mathbf{Z}=\mathbf{z }%
\vert\mathcal{F}, \mathcal{Z} ) =2^{-I}$ for each
$\mathbf{z}\in\mathcal{Z}$. If $T=t ( \mathbf{Z},\mathbf{R}
) $
is a test statistic, then its distribution in a randomized paired experiment
under the null hypothesis of no effect is its permutation distribution, that
is, $\Pr( T\geq t \vert\mathcal{F}, \mathcal{Z} ) =\Pr( t ( \mathbf
{Z},\mathbf{R}%
) \geq t \vert\mathcal{F}, \mathcal{Z} )
=\Pr( t ( \mathbf{Z},\mathbf{r}_{C} ) \geq
t \vert\mathcal{F}, \mathcal{Z} ) $
equals$ \llvert\{ \mathbf{z}\in\mathcal{Z}\dvtx t ( \mathbf{Z},\mathbf
{r}_{C} ) \geq t \} \rrvert/\llvert\mathcal
{Z}%
\rrvert$, because, under $H_{0}$, $\mathbf{R}=\mathbf{r}_{C}$ is
fixed by
conditioning on $\mathcal{F}$, and $\mathbf{Z}$ is uniform on~$\mathcal{Z}$.

The treated-minus-control pair difference in observed responses in pair $i$
is
\[
Y_{i}= ( Z_{i1}-Z_{i2} ) ( R_{i1}-R_{i2}
) =Z_{i1} ( r_{Ti1}-r_{Ci2} ) +Z_{i2} (
r_{Ti2}-r_{Ci1} ),
\]
which equals $ ( Z_{i1}-Z_{i2} ) ( r_{Ci1}-r_{Ci2} )
=\pm( r_{Ci1}-r_{Ci2} ) $ if $H_{0}$ is true. Figure~\ref{fig3} depicts
the pair differences in 2006 test scores, $Y_{i}$. In general, $Y_{i}%
=Z_{i1} ( r_{Ti1}-r_{Ci2} ) +Z_{i2} ( r_{Ti2}-r_{Ci1}%
) $, which equals $Y_{i}=\tau+\varepsilon_{i}$ with $\varepsilon
_{i}= (
Z_{i1}-Z_{i2} ) ( r_{Ci1}-r_{Ci2} ) $ if the treatment
effect is a constant shift, $r_{Tij}-r_{Cij}=\tau,\ \forall i,j$. Let
$q_{i}\geq0$ be a function of $\llvert Y_{1}\rrvert,\ldots,\llvert
Y_{I}\rrvert$ such that $q_{i}=0$ if $\llvert Y_{i}\rrvert=0$.
Let $\operatorname{sgn} ( y ) =1$ if $y>0$ and $\operatorname{sgn} (
y ) =0$ if $y\leq0$. A general signed rank statistic is of the form
$T=\sum_{i=1}^{I}\operatorname{sgn} ( Y_{i} ) q_{i}$. In a paired,
randomized experiment under $H_{0}$, the null distribution $\Pr(
T\geq t \vert\mathcal{F}, \mathcal{Z} ) $ of
$T$ is
the distribution of the sum of $I$ independent random variables taking the
values $q_{i}$ or 0 each with probability $1/2$ if $\llvert Y_{i}%
\rrvert>0$ and the value $0$ with probability 1 if $\llvert
Y_{i}\rrvert=0$. For instance, if $q_{i}$ is the rank of $\llvert
Y_{i}\rrvert$, this yields the usual null distribution of Wilcoxon's
signed-rank statistic.

For certain rank statistics, such as Wilcoxon's statistic, the expectation
$\mu$ of the test statistic under the null hypothesis $H_{0}$, namely,
$\mu=\mathrm{E} \{ t ( \mathbf{Z},\mathbf{r}_{C} )
\vert\mathcal{F}, \mathcal{Z} \} $, does not depend
upon $\mathbf{r}_{C}$, and in these cases \citet{HodLeh63} proposed
estimating a constant shift effect $\tau$ by $\hat{\tau}$ that solves
$t ( \mathbf{Z},\mathbf{R}-\mathbf{Z}\hat{\tau} ) \doteq\mu$.

\subsection{Sensitivity analysis}\label{ssSenAnalysis}

A simple model for sensitivity analysis in paired observational studies
[\citet{Ros87}] has a sensitivity parameter $\Gamma\geq1$ and
asserts that
$\Pr( \mathbf{Z}=\mathbf{z }\vert\mathcal{F}, \mathcal{Z} ) =%
{\prod_{i=1}^{I}}
\pi_{i}^{z_{i}} ( 1-\pi_{i} ) ^{1-z_{i}}$ for $\mathbf{z}%
\in\mathcal{Z}$ where $1/ ( 1+\Gamma) \leq\pi_{i}\leq
\Gamma/ ( 1+\Gamma) $ for each $i$ but $\pi_{i}$ is otherwise
unknown. When $\Gamma=1$, the distribution of treatment assignments
is the
randomization distribution, $\Pr( \mathbf{Z}=\mathbf{z }%
\vert\mathcal{F}, \mathcal{Z} ) =2^{-I}$, but when
$\Gamma>1$
the distribution of treatment assignments $\Pr( \mathbf{Z}%
=\mathbf{z }\vert\mathcal{F}, \mathcal{Z} ) $ is
unknown to a
degree bounded by $\Gamma$. Therefore, when $\Gamma=1$ conventional
randomization inferences are obtained, for instance, randomization tests,
confidence intervals formed by inverting randomization tests [e.g., \citet{Mar79}] and \citet{HodLeh63}
point estimates. For $\Gamma>1$, one
obtains instead an interval of $P$-values, an interval of point
estimates or
an interval of endpoints for a confidence interval, the interval becoming
longer as $\Gamma$ increases. One asks: how large must $\Gamma$ be,
how far
must the observational study deviate from a randomized experiment,
before the
range of inferences becomes uninformative? For instance, how large must
$\Gamma$ be before the interval of $P$-values includes values above and below
$\alpha$, conventionally $\alpha=0.05$? This model may be expressed
explicitly in terms of the unobserved covariate $u_{ij}$, derived from more
basic assumptions similar to those in \citet{Coretal59}, and easily
extended to matching with multiple controls, full matching, unmatched
comparisons, covariance adjustment of matched pairs, etc.; see \citet{Ros02}, Section~4;
(\citeyear{Ros07}). Although the sensitivity analysis permits
the unobserved
covariate $u_{ij}$ to vary from student to student, there is nothing to
prevent $u_{ij}$ from being constant for children from the same family
or the
same social clique, so $u_{ij}$ can represent some unmeasured form of
clustering. For other models for sensitivity analysis in observational
studies, see \citet{Gas92}, \citet{HosHanHol10}, \citet{Mar97},
\citet{RosRub83}, \citet{Sma07}, \citet{Yan84} and \citet{YuGas05}.

For a specific $\Gamma\geq1$, define $\overline{\overline{T}}$ as the
sum of
$I$ independent random variables taking the value $q_{i}$ with probability
$\Gamma/ ( 1+\Gamma) $ and the value 0 with probability
$1/ (
1+\Gamma) $, and define $\overline{T}$ similarly but with
$\Gamma/ ( 1+\Gamma) $ and $1/ ( 1+\Gamma) $
interchanged. In the presence of a bias of magnitude $\Gamma$, the null
distribution of $T$ under $H_{0}$ is unknown, but it is easily shown to be
bounded by two-known distributions,
%
%
\begin{equation}\label{eqSenBound}
\Pr( \overline{T}\geq t \vert\mathcal{F}, \mathcal{Z} ) \leq\Pr(
T\geq t \vert\mathcal{F}, \mathcal{Z} ) \leq\Pr( \overline{\overline
{T}}\geq t
\vert\mathcal{F}, \mathcal{Z} )\qquad\mbox{for all }t;
\end{equation}
see Rosenbaum (\citeyear{Ros87}; \citeyear{Ros02}, Section~4). For reasonable scores,
$q_{i}$, the bounds
in~(\ref{eqSenBound}) may be approximated as $I\rightarrow\infty$ using the
central limit theorem:
%
%
\begin{eqnarray}\label{eqSenBoundApprox}
\hspace*{-50pt}\Pr( \overline{\overline{T}}\geq t_{\Gamma,\alpha}%
\vert\mathcal{F},
\mathcal{Z} ) \approx\alpha
\nonumber\\[2pt]\\[-20pt]
\eqntext{\displaystyle\mbox{for }t_{\Gamma,\alpha}=\frac
{\Gamma}{1+\Gamma}\sum_{i=1}^{I}q_{i}+\Phi
^{-1} ( 1-\alpha) \sqrt{\frac{\Gamma}{ ( 1+\Gamma)
^{2}}\sum _{i=1}^{I}q_{i}^{2}},}
\end{eqnarray}
where $\Phi( \cdot) $ is the standard Normal cumulative
distribution, so that, if $T\geq t_{\Gamma,\alpha}$, then the
approximation to
the maximum one-sided $P$-value is at most $\alpha$ when the sensitivity
analysis allows for an unmeasured bias of at most $\Gamma$. For
instance, if
$T\geq t_{1.25, 0.05}$, then the entire interval of possible one-sided $P
$-values obtained from a bias of $\Gamma=1.25$ is below $\alpha=0.05$,
and a
bias of magnitude $\Gamma=1.25$ is too small to explain the observed
value of
the test statistic $T$.

For statistics such as Wilcoxon's statistic, the sum $\sum
_{i=1}^{I}q_{i}$ in
(\ref{eqSenBoundApprox}) does not depend upon $\mathbf{r}_{C}$, and the
expectation of $T$ under $H_{0}$ is bounded by the expectations of
$\overline{T}$ and $\overline{\overline{T}}$, namely, $\overline{\mu
}_{\Gamma
}= ( 1+\Gamma) ^{-1}\sum_{i=1}^{I}q_{i}$ and $\overline{\overline{\mu
}}_{\Gamma}= \{ \Gamma/ ( 1+\Gamma) \}
\sum_{i=1}^{I}q_{i}$. In these cases, the interval of possible
Hodges--Lehmann point estimates of a constant shift effect $\tau$ is obtained
by solving $t ( \mathbf{Z},\mathbf{R}-\mathbf{Z}\hat{\tau} )
\doteq\overline{\mu}_{\Gamma}$ and $t ( \mathbf{Z},\mathbf{R}%
-\mathbf{Z}\hat{\tau} ) \doteq\overline{\overline{\mu}}_{\Gamma}$;
see Rosenbaum (\citeyear{Ros93}; \citeyear{Ros02}, Section~4). This is done in Table~\ref{tabsenhl}
below.
A similar approach may be used with Huber's $M$-estimates including the mean
of the $I$ paired differences; see Rosenbaum (\citeyear{Ros07}, \citeyear{Ros13}) and
Section~\ref{ssMeanMstatistic}.

\subsection{Power of a sensitivity analysis and design sensitivity;
testing one hypothesis twice}\label{ssDesignSen}

If there was no bias from an unmeasured covariate $u_{ij}$ and if the
treatment had an effect so $H_{0}$ is false, then we could not be
certain of
this from the observed data, and the best we could hope to say is that the
conclusions are insensitive to a moderately large bias $\Gamma$, for instance,
that $T\geq t_{\Gamma,\alpha}$ for a moderately large $\Gamma$. The
power of
a one-sided, $\alpha$-level sensitivity analysis at a specific $\Gamma$
is the
probability that we will be able to say this, that is, the power is the
probability that $T\geq t_{\Gamma,\alpha}$ when there actually is no bias,
$\Pr( \mathbf{Z}=\mathbf{z }\vert\mathcal{F}, \mathcal{Z} ) =2^{-I}$,
and the $Y_{i}$ are generated by some model
with a treatment effect, such as $Y_{i}\sim_{\mathrm{i.i.d.}}N (
\tau,1 ) $; see Rosenbaum (\citeyear{Ros04}; \citeyear{Ros10N1}, Part~III). When $\Gamma=1$,
the power of a sensitivity analysis is the same as the power of a
randomization test.

Under mild conditions, for a given model such as $Y_{i}\sim_{\mathrm
{i.i.d.}}N ( \tau,1 ) $ and a given statistic $T$ such as Wilcoxon's
statistic, there is a value $\widetilde{\Gamma}$ called the design sensitivity
such that, as the sample size increases, $I\rightarrow\infty$, the
power of
the sensitivity analysis tends to 1 when the analysis is performed with
$\Gamma<\widetilde{\Gamma}$ and the power tends to 0 with $\Gamma
>\widetilde{\Gamma}$. In words, in this sampling situation with this
statistic, the study will eventually be insensitive to all biases
smaller than
$\widetilde{\Gamma}$ but not to some biases larger than $\widetilde
{\Gamma}$.
Just as the power of a randomization test is affected by the choice
of test
statistic, so too is the power of a sensitivity analysis and the design
sensitivity affected by the choice of test statistic. For instance, if
$Y_{i}\sim_{\mathrm{i.i.d.}}N ( \tau,1 ) $, then with $\tau=1/2$, the
design sensitivity is $\widetilde{\Gamma}=3.2$ for Wilcoxon's signed-rank
statistic and $\widetilde{\Gamma}=3.6$ for Brown's (\citeyear{Bro81}) combined quantile
average, so at $\Gamma=3.4$, the power of Wilcoxon's statistic is
tending to 0
as $I\rightarrow\infty$ while the power of Brown's statistic is tending
to 1;
see \citet{Ros10N2}.

Better design sensitivities are possible with other statistics. In \citet{Ros11}, a $U$-statistic named $ ( m,\underline{m},\overline{m} ) $
with $1\leq\underline{m}\leq\overline{m}\leq m<I$ is defined by looking
at all
subsets of $m$ of the $Y_{i}$, sorting these $m$ observations into increasing
order by $\llvert Y_{i}\rrvert$, counting the number of positive
$Y_{i}$ among those in positions $\underline{m},\underline{ m}+1,\ldots,\overline{ m}$ in this order, and averaging over the ${I\choose m}$ subsets
of size $m$; it is a signed-rank statistic with $q_{i}={I\choose m}^{-1}
\sum_{\ell=\underline{m}}^{\overline{m}}{a_{i}-1\choose \ell-1}
{I-a_{i}\choose m-\ell}$, where $a_{i}$ is the rank of $\llvert Y_{i}\rrvert$ and
${A\choose B}$ is defined to equal 0 for $B<0$. In particular, $ (
m,\underline{m},\overline{m} ) = ( 1,1,1 ) $ is the sign test
statistic, $ ( m,\underline{m},\overline{m} ) = (
2,2,2 ) $ is the $U$-statistic that closely approximates Wilcoxon's
signed-rank statistic [\citet{Leh75}], and $ ( m,\underline
{m},\overline{m} ) = ( m,m,m ) $ is Stephenson's (\citeyear{Ste81}) statistic. If
$Y_{i}=\tau+\varepsilon_{i}$ with $\tau=1/2$ and $\varepsilon_{i}\sim
_{\mathrm
{i.i.d.}}N ( 0,1 ) $, then Wilcoxon's test $ ( m,\underline{m},\overline
{m} ) =
( 2,2,2 ) $ has $\widetilde{\Gamma}=3.2$ as
before, while $\widetilde{\Gamma}=5.5$ for $ ( m,\underline{m},\overline
{m} ) = ( 20,16,20 ) $, $\widetilde{\Gamma}=6.9$
for $ ( m,\underline{m},\overline{m} ) = ( 20,18,20 ) $,
and $\widetilde{\Gamma}=10.1$ for $ ( m,\underline{m},\overline
{m} ) = ( 20,20,20 ) $. If $Y_{i}=\tau+\varepsilon_{i}$ with
$\tau=1$ and the $\varepsilon_{i}$ are independently distributed with a
$t$-distribution on 4 degrees of freedom, then Wilcoxon's test $ (
m,\underline{m},\overline{m} ) = ( 2,2,2 ) $ has
$\widetilde{\Gamma}=6.8$, while $\widetilde{\Gamma}=9.4$ for $ (
m,\underline{m},\overline{m} ) = ( 20,16,20 ) $,
$\widetilde{\Gamma}=8.9$ for $ ( m,\underline{m},\overline{m} )
= ( 20,18,20 ) $, and $\widetilde{\Gamma}=7.3$ for $ (
m,\underline{m},\overline{m} ) = ( 20,20,20 ) $. Notably,
Wilcoxon's statistic has relatively poor performance in all these situations,
while the best test statistic depends upon the tails of the
distribution of
$\varepsilon_{i}$.

\begin{figure}

\includegraphics{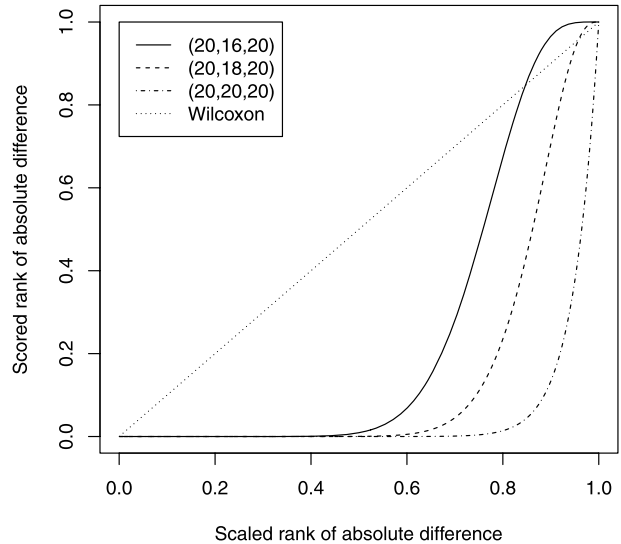}

\caption{Four ways of scaling the ranks of absolute difference $|Y_{i}|$
in post-treatment test scores.}\label{fig4}
\end{figure}

Figure~\ref{fig4} shows $q_{i}/\max q_{j}$ against Wilcoxon's ranks $a_{i}/\max a_{j}$
for Wilcoxon's statistic $ ( m,\underline{m},\overline{m} )
= ( 2,2,2 ) $ and for $ ( m,\underline{m},\overline
{m} )
= ( 20,16,20 ) $, $ ( 20,18,20 ) $ and $ (
20,20,20 ) $. Unlike Wilcoxon's statistic, the other three statistics
largely ignore $Y_{i}$ with small $\llvert Y_{i}\rrvert$, but
do this
in varying degrees. As discussed in \citet{Ros10N2}, reduced
attention to
$Y_{i}$ with small $\llvert Y_{i}\rrvert$ tends to increase design
sensitivity, $\widetilde{\Gamma}$, and this explains, for example, the
superior design sensitivity of Brown's (\citeyear{Bro81}) statistic when compared to
Wilcoxon's statistic.

In \citet{Ros12N2}, several tests are performed of the same null hypothesis
$H_{0}$ using different test statistics, and the smallest upper bound
on the
$P$-value from these several tests is corrected for multiple testing, an
appropriate correction being quite small because of the strong dependence
between several tests of the same null hypothesis using the same data.
The
correction approximates the joint distribution of the upper bound statistics
by a multivariate Normal distribution. This combined procedure
achieves the
best design sensitivity of the several component tests; for example, using
$ ( m,\underline{m},\overline{m} ) = ( 20,16,20 ) $,
$ ( 20,18,20 ) $ and $ ( 20,20,20 ) $ jointly, the
combination would have $\widetilde{\Gamma}=10.1$ for the Normal distribution
above and $\widetilde{\Gamma}=9.4$ for the $t$-distribution above, having
selected the best test for each distribution. This procedure is used in
Section~\ref{ssTestTwice} for the study in Section~\ref{ssIntroExample}.

\subsection{Reducing heterogeneity reduces sensitivity to unmeasured
biases}\label{ssHeterogeneity}

As mentioned in Section~\ref{ssIntroPlan}, reducing heterogeneity tends
to reduce
sensitivity to unmeasured biases. For instance, if $Y_{i}=\tau
+\varepsilon_{i}$
with $\tau=1/2$ and $\varepsilon_{i}\sim_{\mathrm{i.i.d.}}N ( 0,\sigma
^{2} ) $, then Wilcoxon's signed-rank statistic has design sensitivity
$\widetilde{\Gamma}=3.2$ as before if $\sigma=1$, but it has design
sensitivity $\widetilde{\Gamma}=11.7$ if the standard deviation is cut in
half, $\sigma=1/2$. Similarly, in this sampling situation, the $U$-statistic
$ ( m,\underline{m},\overline{m} ) = ( 20,18,20 ) $
has design sensitivity $\widetilde{\Gamma}=6.9$ as before if $\sigma
=1$, but it
has design sensitivity $\widetilde{\Gamma}=91.6$ if the standard
deviation is
cut in half, $\sigma=1/2$. This phenomenon is not tied to Normal
distributions or to particular test statistics, and it is discussed in detail
in \citet{Ros05}. As discussed there, reducing heterogeneity
$\sigma$
confers benefits for sensitivity analyses that cannot be produced by
increasing the sample size, $I$, because these benefits occur even in the
limit as $I\rightarrow\infty$. The hope in Section~\ref
{ssStep2optimalpairing} is
that the reduction in dispersion of $Y_{i}$ seen in Figure~\ref{fig3} may yield reduced
sensitivity to unmeasured biases. As just seen, reducing the scale
$\sigma$
by half has a large effect on design sensitivity, $\widetilde{\Gamma}$, but
the reduction in Figure~\ref{fig3} is closer to 15\% than to 50\%. Again,
Section~\ref{eqCardMatch} achieved a reduction in heterogeneity of the $Y_{i}$
without altering their mean, $I^{-1}\sum Y_{i}$, by balancing covariates
$\mathbf{x}$ first using (\ref{eqCardMatch}), then pairing students for
pretreatment 2004 test scores that predict posttreatment 2006 test scores.

\subsection{Amplification: 2-dimensional interpretation of a
1-dimensional sensitivity analysis}

\label{ssAmplify}

For analysis and reporting, it is convenient to have a one-dimensional
sensitivity analysis defined in terms of a single parameter, $\Gamma$.
At
$\Gamma=1$ the distribution of treatment assignments is randomized, but as
$\Gamma\rightarrow\infty$ any treatment assignment probabilities $\pi_{i}$
become possible, so $\Gamma$ is a way of indexing the magnitude of departure
from random assignment, not a device for giving that departure a specific
form. The parameter $\Gamma$ measures the impact of the unobserved covariate
$u_{ij}$ on the treatment assignment probabilities $\pi_{i}$, placing no
restriction on the relationship between $u_{ij}$ and the outcome
$Y_{i}$, so
$u_{i1}-u_{i2}$ may be strongly related to $Y_{i}$ under $H_{0}$. For
interpretation, it is sometimes convenient to reexpress this one
analysis in
terms of $\Gamma$ instead as an equivalent two-dimensional analysis
with a
parameter $\Lambda$ that controls the relationship between
$u_{i1}-u_{i2}$ and
treatment assignment $Z_{i1}-Z_{i2}=\pm1$ and another parameter $\Delta
$ that
controls the relationship under $H_{0}$ between $u_{i1}-u_{i2}$ and the sign
of $Y_{i}$. Under $H_{0}$, $\Lambda=2$ means that an imbalance in $u$ at
most doubles the odds of treatment, $Z_{i1}-Z_{i2}=1$, while $\Delta=2$ means
that $u$ at most doubles the odds of a positive response difference,
$Y_{i}%
>0$, and the parameter $\Delta$ is defined in terms of Wolfe's (\citeyear{Wol74})
semiparametric family of deformations of a distribution symmetric about zero;
see \citet{RosSil09} for technical specifics where $\Gamma
= (
\Lambda\Delta+1 ) / ( \Lambda+\Delta) $. Such a map of
each value of one sensitivity parameter $\Gamma$ into an exactly equivalent
curve $\Gamma= ( \Lambda\Delta+1 ) / ( \Lambda+\Delta)
$ of a two-parameter $ ( \Lambda,\Delta) $ sensitivity analysis
is called an amplification. For instance, the curve corresponding with
$\Gamma=1.5$ includes $ ( \Lambda,\Delta) = ( 2,4 ) $
as $1.5= ( 2\times4+1 ) / ( 2+4 ) $, but it also
includes $ ( \Lambda,\Delta) = ( 4,2 ) $ and also
$ ( \Lambda,\Delta) = ( 2.5,2.75 ) $. That is, under
$H_{0}$, $\Gamma=1.5$ is equivalent to an unobserved covariate $u$ that
doubles the odds of treatment, $\Lambda=2$, and quadruples the odds of a
positive response difference $Y_{i}>0$, $\Delta=4$, and is also
equivalent to an
analysis in which $u$ quadruples the odds of treatment, $\Lambda=4$, and
doubles the odds of a positive response difference, $\Delta=2$.

\section{Sensitivity analysis in a cardinality match paired for
heterogeneity}\label{secSenInExample}

\subsection{Analyses using one rank statistic}\label{ssOneRankStatistic}

Using the methods in Sections\break \ref{ssSenAnalysis}~and~\ref{ssDesignSen}, Table
\ref{tabsenPval} examines the sensitivity of the null hypothesis $H_{0}$ of no treatment effect
in the two pairings in Section~\ref{ssStep2optimalpairing} of the same
cardinality
match in Section~\ref{ssStep1CardinalityMatch}. The table also uses
two test
statistics from Section~\ref{ssDesignSen}, namely, the Wilcoxon
statistic with
$ ( m,\underline{m},\overline{m} ) = ( 2,2,2 ) $ and one
version of the $U$-statistic with $ ( m,\underline{m},\overline
{m} )
= ( 20,18,20 ) $. Table~\ref{tabsenPval}
records the upper bound on the one-sided $P$-value testing $H_{0}$, so the
comparison is insensitive to a bias of $\Gamma$ if this upper bound is less
than the conventional $\alpha=0.05$. Notably in Table
\ref{tabsenPval}, Wilcoxon's statistic with pairing based on all covariates becomes sensitive
between $\Gamma=1.3$ and $\Gamma=1.4$, whereas the $U$-statistic with pairing
based on four pretreatment test scores becomes sensitive between $\Gamma=1.6$
and $\Gamma=1.7$. Looking at the row $\Gamma=1.4$ in Table~\ref{tabsenPval}
suggests that in this one example, the choice of pairing and the choice of
test statistic are comparable in importance but separate effects.%

\begin{table}[t]
\tabcolsep=10pt
\tablewidth=258pt
\caption{Upper bounds on the one-sided $P$-value testing the null
hypothesis $H_{0}$ of no
treatment effect, using either Wilcoxon's statistic or one version of
the $U$-statistic, with pairing based
either on all covariates or just the four pretreatment test scores. The
$Y_{i}$ are less heterogeneous
when the pairing controlled just the four pretreatment test scores}\label{tabsenPval}
\begin{tabular*}{\tablewidth}{@{\extracolsep{\fill}}@{}lcccc@{}}
\hline
& \multicolumn{4}{c@{}}{\textbf{Covariates used in pairing}}\\[-6pt]
& \multicolumn{4}{c@{}}{\hrulefill}\\
& \multicolumn{2}{c}{\textbf{Wilcoxon statistic $\bolds{(2,2,2)}$}} & \multicolumn{2}{c@{}}{\textbf{$\bolds{U}$-statistic} $\bolds{(20,18,20)}$}\\[-6pt]
& \multicolumn{2}{c}{\hrulefill} & \multicolumn{2}{c@{}}{\hrulefill}\\
$\bolds{\Gamma}$ & \textbf{All} & \textbf{4 test scores} & \textbf{All} & \textbf{4 test scores} \\
\hline
1 & 0.0000 & 0.0000 & 0.0000 & 0.0000 \\
1.1 & 0.0000 & 0.0000 & 0.0000 & 0.0000 \\
1.2 & 0.0001 & 0.0000 & 0.0005 & 0.0000 \\
1.3 & 0.0131 & 0.0008 & 0.0062 & 0.0001 \\
1.4 & 0.1986 & 0.0367 & 0.0378 & 0.0010 \\
1.5 & 0.6681 & 0.3031 & 0.1341 & 0.0078 \\
1.6 & 0.9488 & 0.7506 & 0.3149 & 0.0356 \\
1.7 & 0.9971 & 0.9638 & 0.5418 & 0.1099 \\
\hline
\end{tabular*}
\end{table}

%
%

Table~\ref{tabsenhl}
is similar in structure to Table
\ref{tabsenPval}%
, but it reports the minimum Hodges--Lehmann point estimate $\hat{\tau}$ of
an additive treatment effect $\tau$ from Section~\ref{ssSenAnalysis}. For
$\Gamma=1$, the interval is a single point, and in Table
\ref{tabsenPval}
is not far from the mean of the $Y_{i}$, namely, 17.5 points on the
total of
mathematics and language tests, as depicted in Figure~\ref{fig3}. At $\Gamma=1.7$,
the minimum estimate from Wilcoxon's test applied to pairs matched for all
covariates is $-6.9$, so not-for-profit schools could be harmful, but at
$\Gamma=1.7$ the minimum estimate from the $U$-statistic applied to pairs
matched for the four pretreatment test scores is still positive 3.2. %

%
%
\begin{table}[b]
\tabcolsep=12pt
\tablewidth=258pt
\caption{Minimum Hodges--Lehmann point estimate of an additive effect
$\tau$ of attending a not-for-profit school rather than a for-profit school,
using either Wilcoxon's statistic
or one version of the $U$-statistic, with pairing based
either on all covariates or just the four pretreatment test scores}\label{tabsenhl}
\begin{tabular*}{\tablewidth}{@{\extracolsep{\fill}}@{}ld{3.1}d{2.1}d{2.1}d{2.1}@{}}
\hline
& \multicolumn{4}{c@{}}{\textbf{Covariates used in pairing}}\\[-6pt]
& \multicolumn{4}{c@{}}{\hrulefill}\\
& \multicolumn{2}{c}{\textbf{Wilcoxon statistic $\bolds{(2,2,2)}$}} & \multicolumn{2}{c@{}}{\textbf{$\bolds{U}$-statistic} $\bolds{(20,18,20)}$}\\[-6pt]
& \multicolumn{2}{c}{\hrulefill} & \multicolumn{2}{c@{}}{\hrulefill}\\
$\bolds{\Gamma}$ & \multicolumn{1}{c}{\textbf{All}} & \multicolumn{1}{c}{\textbf{4 test scores}} & \multicolumn{1}{c}{\textbf{All}} & \multicolumn{1}{c@{}}{\textbf{4 test scores}} \\
\hline
1 & 17.9 & 17.1 & 14.8 & 16.9 \\
1.1 & 13.4 & 13.3 & 12.1 & 14.4 \\
1.2 & 9.4 & 9.9 & 9.5 & 12.1 \\
1.3 & 5.6 & 6.7 & 7.2 & 10.1 \\
1.4 & 2.1 & 3.8 & 5.1 & 8.1 \\
1.5 & -1.1 & 1.1 & 3.1 & 6.4 \\
1.6 & -4.1 & -1.4 & 1.3 & 4.7 \\
1.7 & -6.9 & -3.8 & -0.3 & 3.2 \\
\hline
\end{tabular*}
\end{table}

In brief, in terms of significance levels testing no effect or point estimates
$\hat{\tau}$ of the magnitude of effect, results are less sensitive to
unmeasured biases using a pairing that stabilizes $Y_{i}$ and a test statistic
that largely ignores $Y_{i}$ with small $\llvert Y_{i}\rrvert$.

\subsection{Analyses using the mean or one $M$-statistic}\label
{ssMeanMstatistic}

The analyses in Section~\ref{ssOneRankStatistic} used rank statistics,
such as
Wilcoxon's signed-rank statistic, but an alternative is to use the mean
or one
of Huber's $M$-statistics. There is a parallel sensitivity analysis
for the
mean of the 1907 treated-minus-control pair differences or for other
$M$-statistics computed from these differences; see \citet{Ros07}. The
permutational $t$-test [\citet{Wel37}] is essentially the same as a signed-rank
statistic with $q_{i}=\llvert Y_{i}\rrvert$ and Maritz's (\citeyear{Mar79})
permutational $M$-statistic essentially uses a different definition of $q_{i}
$, so that the sensitivity analysis is similar to Section~\ref{ssSenAnalysis};
again, see \citet{Ros07} for some necessary but omitted details. For both
re-pairings, the sample mean difference is 17.5 points, as in Figure~\ref{fig3},
and it
would be unbiased for the average treatment effect if $\Gamma=1$. In the
absence of bias, $\Gamma=1$, the permutational $t$-test rejects the null
hypothesis of no effect with one-sided \mbox{$P$-}value $4.3\times10^{-13}$ when
pairing with all covariates and with $P$-value $1.1\times10^{-16}$ when
pairing for the four baseline test scores. At $\Gamma=1.4$, the upper bound
on the \mbox{$P$-}value from the permutational $t$-test is 0.098 when pairing
for all
covariates and is 0.005 when pairing for the four baseline test scores.
When
pairing for the four test scores, the upper bound on the $P$-value from the
permutational \mbox{$t$-}test is 0.082 at $\Gamma=1.5$, but the smallest possible
point estimate of the mean effect of the treatment is still 3 points.

As in the case of rank statistics, reducing the weight attached to $Y_{i}$
with small $\llvert Y_{i}\rrvert$ increases the design
sensitivity of
$M$-statistics; see \citet{Ros13}. One such $M$-test combines Huber's
outer trimming with some inner trimming: specifically, (i) it gives zero
weight to $Y_{i}$ with $\llvert Y_{i}\rrvert$ less than half the
median of the $\llvert Y_{i}\rrvert$, (ii) it gives constant
weight of
1 to $Y_{i}$ greater than three times the median of the $\llvert
Y_{i}\rrvert$, and (iii) it rises linearly from 0 to 1 between
half the
median of the $\llvert Y_{i}\rrvert$ and three times the median
of the
$\llvert Y_{i}\rrvert$. As anticipated from calculations of its
design sensitivity in \citet{Ros13}, this statistic reports somewhat less
sensitivity to unmeasured bias than does the permutational $t$-test: at
$\Gamma=1.5$, the upper bound on the $P$-value is 0.032 when pairing
for the
four test scores.

In brief, the patterns seen in Section~\ref{ssOneRankStatistic} for
rank statistics
also occur for the mean and for $M$-statistics. For all of these statistics,
reducing heterogeneity of $Y_{i}$ by re-pairing for a few key covariates
results in reduced sensitivity to unmeasured biases.

\subsection{Analyses that use several test statistics to test the same
hypothesis}\label{ssTestTwice}%

%
%
\begin{table}
\tabcolsep=0pt
\tablewidth=250pt
\caption{Sensitivity analysis for two ways of pairing the same 3814
students into 1907 pairs. Upper bound on the one-sided $P$-value for several values of
$\Gamma$. When pairing for all covariates, the bound is 0.0498 at $\Gamma
=1.42$. When pairing
for the 4 baseline test scores, the bound is 0.0491 at $\Gamma=1.77$}\label{tabsen}
\begin{tabular*}{\tablewidth}{@{\extracolsep{\fill}}@{}lcc@{}}
\hline
& \multicolumn{2}{c@{}}{\textbf{Pairing of 3814 students}}\\[-6pt]
& \multicolumn{2}{c@{}}{\hrulefill}\\
$\bolds{\Gamma}$ & \textbf{For all covariates} & \textbf{For 4 baseline scores}\\
\hline
1 & 0.0000 & 0.0000 \\
1.1 & 0.0000 & 0.0000 \\
1.2 & 0.0001 & 0.0000 \\
1.3 & 0.0034 & 0.0001 \\
1.4 & 0.0364 & 0.0006 \\
1.5 & 0.1011 & 0.0028 \\
1.6 & 0.2004 & 0.0101 \\
1.7 & 0.3333 & 0.0275 \\
1.75 & 0.4074 & 0.0421 \\
\hline
\end{tabular*}
\end{table}

Table~\ref{tabsen}
uses three test statistics to test the one null hypothesis $H_{0}$ of no
treatment effect, correcting for multiple testing, as discussed in
Section~\ref{ssDesignSen} and \citet{Ros12N2}. Specifically, the
test uses the
$U$-statistics with $ ( m,\underline{m},\overline{m} ) = (
20,16,20 ) $, $ ( 20,18,20 ) $ and $ ( 20,20,20 )
$. With short-tailed distributions like the Normal, $ (
20,20,20 ) $ is the best of these three in terms of design sensitivity
$\widetilde{\Gamma}$, but with the slightly thicker tails of a $t$%
-distribution on 4 degrees of freedom, $ ( 20,16,20 ) $ is best.
Table
\ref{tabsen}
reports the smallest of the three upper bounds on $P$-values after correcting
for testing three times, the appropriate correction being small because
of the
strong positive dependence between three tests of the same hypothesis
based on
the same data.

As theory anticipates, Table
\ref{tabsen}
reports somewhat less sensitivity to unmeasured bias than the fixed
choices of
test statistic in Table
\ref{tabsenPval}%
. As in Table
\ref{tabsenPval}%
, the less heterogeneous pairing based on four pretreatment test scores yields
less sensitivity to unmeasured bias than pairing for all covariates.

\begin{figure}

\includegraphics{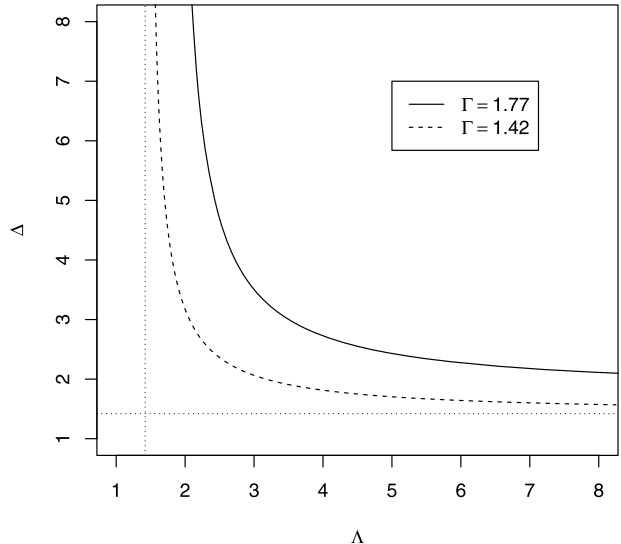}

\caption{Amplification or re-expression of a sensitivity involving one
parameter $\Gamma$ at $\Gamma=1.77$ or $\Gamma=1.42$ into an equivalent
sensitivity analysis involving two parameters. Here, $\Lambda$ controls the
relationship between treatment assignment, namely $Z_{i1}-Z_{i2}$,
and the unobserved covariate, $u_{i1}-u_{i2}$, and $\Delta$ controls
the relationship between a positive response difference under
${H}_{0}$, namely $R_{i1}-R_{i2} = r_{Ci1}-r_{Ci2}$,
and the unobserved covariate, $u_{i1}-u_{i2}$. For instance,
$(\Lambda,\Delta) = (3.00, 2.06)$ is equivalent to $\Gamma=1.42$ while
$(\Lambda,\Delta) = (3.00, 3.50)$ is equivalent to $\Gamma=1.77$. The
dotted lines are at the asymptote of 1.42 for $\Gamma=1.42$.}\label{fig5}
\end{figure}

Figure~\ref{fig5} depicts the amplification of the sensitivity analysis in Table
\ref{tabsen}%
, so that, as in Section~\ref{ssAmplify}, the single values of $\Gamma
=1.42$ and
$\Gamma=1.77$ are expressed as the corresponding curves of $ (
\Lambda,\Delta) $ at $\Gamma= ( \Lambda\Delta+1 ) / (
\Lambda+\Delta) $. In particular, the curve for $\Gamma=1.42$
includes $ ( \Lambda,\Delta) = ( 3, 2.06 ) $, or an
unobserved covariate $u$ that roughly a triples the odds of treatment and
doubles the odds of a positive difference in test scores. In
contrast, the
$\Gamma=1.77$ includes $ ( \Lambda,\Delta) = (
3, 3.50 ) $, or roughly a tripling of the odds of treatment and a
3.5-fold increase in the odds of a positive difference in test scores.
The
reduction in heterogeneity in Figure~\ref{fig3} moves the degree of sensitivity from
$\Gamma=1.42$ to $\Gamma=1.77$, and for $\Lambda=3$ this is a move from
$\Delta\doteq2$ to $\Delta\doteq3.5$. In view of this, a meaningful
reduction in sensitivity to unmeasured biases was produced by balancing all
covariates first in Section~\ref{ssStep1CardinalityMatch} and closely
pairing for
the predictive covariates in Section~\ref{ssStep2optimalpairing}.

\section{Summary}

In matching, covariate balance refers to the distributions of the observed
covariate $\mathbf{x}$ in treated and control groups. Cardinality matching
constructs the largest matched sample that satisfies specified constraints
(\ref{eqOneConstraint}) on covariate balance $\mathbf{x}$, ignoring who is
paired with whom. With this first task accomplished, with comparable groups
in hand, the pairing can then emphasize a subset of covariates expected to
predict the outcome and hence to reduce heterogeneity of the
treated-minus-control pair differences $Y$. In the example, one
pairing used
all observed covariates, the other used only pretreatment test scores, with
precisely the same students in both pairings, differing only in who was paired
with whom. The same size treatment effect with less heterogeneity or
dispersion of $Y$ tends to be less sensitive to unmeasured biases, that is,
reduced heterogeneity increases the design sensitivity $\widetilde
{\Gamma}$;
see Section~\ref{ssHeterogeneity}. In the example, the mean pair
difference in
$Y$ of 17.5 test score points was meaningfully less sensitive to unmeasured
biases when a pairing based on all covariates was replaced by a pairing
focused on a few predictive covariates yielding a modest reduction in
heterogeneity from a standard deviation of $Y$ of 105.5 to 90.9. As
seen in
the sequence of sensitivity analyses that began with the conventional match
and analysis in the first column of Table~\ref{tabsenPval} and ended with the proposed match
and analysis in the last column of Table~\ref{tabsen}, better matching algorithms that
reduce heterogeneity together with better statistical tests yielded a
substantial reduction in the reported sensitivity to unmeasured biases.
Moreover, as discussed in Section~\ref{secSensitivity}, statistical theory
suggests this reduction in reported sensitivity to bias is expected to occur
when there is an actual treatment effect under simple models for the
generation of the data.

\begin{supplement}
\stitle{Supplement to ``Matching for balance, pairing for heterogeneity in an
observational study of
the effectiveness of for-profit and not-for-profit high schools in Chile''}
\slink[doi]{10.1214/13-AOAS713SUPP} 
\sdatatype{.pdf}
\sfilename{aoas713\_supp.pdf}
\sdescription{In an online supplement we provide additional summary tables for
covariate balance.}
\end{supplement}



\printaddresses

\end{document}